\documentclass[sigconf,nonacm]{acmart}

\usepackage[english]{babel}
\usepackage[normalem]{ulem}

\newcommand{\bid}[1]{\mbox{\bf #1\/}}
\newcommand{\rid}[1]{\mbox{\rm #1\/}}
\newcommand{\id}[1]{\mbox{\it #1\/}}

\renewcommand\footnotetextcopyrightpermission[1]{}
\setcopyright{none}

\begin{document}

\title{Network Programming via Computable Products}

\author{Dennis Volpano}
\affiliation{
  \institution{The Johns Hopkins University Applied Physics Lab}
  \streetaddress{11100 Johns Hopkins Rd}
  \city{Laurel}
  \state{MD}
  \country{U.S.A.}}
\email{Dennis.Volpano@jhuapl.edu}

\begin{abstract}
The User Plane Function (UPF) aims to provide network services in the 3GPP 5G core network.
These services need to be implemented on demand inexpensively with provable properties.
Existing network dataplane programming languages are not up to the task.
A new software paradigm is presented for the UPF.
It is inspired by model checking a concurrent reactive system where conceptually
each component of the system is modeled as an extended finite-state machine and
their product is verified.
We show how such a product can be computed for one example of a UPF and how its state invariants can be inferred,
thereby eliminating the need to formally verify the product separately.
Code can be generated from the product and regenerated on the fly to remain optimal
for the probability distribution of network traffic the UPF must process.
\end{abstract}

\maketitle

\section{Introduction}

The User Plane Function (UPF) in the 3GPP 5G core network \cite{3gppupf} needs new techniques for building
software that implements network functions at the edge quickly, reliably with provable guarantees, and inexpensively.
Provisioning devices will likely be fully automatic and the software can be complex.
Practical network functions are reactive systems responding to inputs based on history and time.
They're not just packet-processing pipelines. 
They have control logic that manages timers, caches and mutable state.

Much work has been done in the design of high-level network programming languages
\cite{michel2021,
foster2011, 
loozhou2012, 
openbox,
voellmy2011,
P42014,
dobrescu2014, 
liu2018,
neves2018,
haoli2020b,
netkat,
temporalnetkat, 
click2000,
balldin20}.
In general, they are either too narrow in scope or lack support for reuse and scalable
proofs about mutable state and timers.
Godefroid observed that model checking a concurrent reactive system conceptually amounts to
modeling each component of the system as an extended finite-state machine and then
verifying the product of all such machines \cite{godefroid2016}.
This idea can be applied to the UPF, instances of
which can be defined as the product of independent
concurrent components represented by finite-state recognizers.
A product can be transformed into branching logic and then implemented
on a specific target platform.
The approach is illustrated for a basic switch function
implemented on an open target platform using Intel's
Data Plane Development Kit (DPDK) \cite{dpdk}.

\section{A basic switch function}

We give four independent concurrent components for a 4-port switch UPF.
It has one uplink port, namely port 1, which is
in a different broadcast domain than ports 2-4.
The components are
\begin{enumerate}
\item $H$ -- (hub) floods a frame to every port except
the port at which it arrived and the uplink port.
\item $B$ -- (bridge) forwards a frame to the port behind which the frame's destination MAC address was learned.
\item $M$ -- learns the ports of MAC addresses.
\item $I$ -- interleaves ingress and egress activity
guaranteeing that every received frame is transmitted.
\end{enumerate}
No component depends on another so all are independent and form reusable building blocks of a switch.
Components are recognizers that run concurrently on a trace.
For example, Table~\ref{trace} shows a trace of our 4-port switch in the presence of the ARP protocol \cite{Stevens2012}.
\begin{table}[h]
\caption{A trace of 4-port switch with uplink port 1}
\small
\begin{tabular}{lllll}
{\em time} & {\em dest address\/} ({\em da}) & {\em src address\/} ({\em sa}) & {\em proto} & {\em location} \\
t & ff:ff:ff:ff:ff:ff & 04:0c:ce:d2:08:6c & arpreq & $\{\rid{2i}\}$  \\
t + 1 & ff:ff:ff:ff:ff:ff & 04:0c:ce:d2:08:6c & arpreq & $\{\rid{3e},\rid{4e}\}$ \\
t + 2 & 04:0c:ce:d2:08:6c & 7c:d1:c3:e8:a4:67 & arpreply & $\{\rid{3i}\}$ \\
t + 3 & 04:0c:ce:d2:08:6c & 7c:d1:c3:e8:a4:67 & arpreply & $\{\rid{2e}\}$
\end{tabular}
\label{trace}
\end{table}
Each port is divided into an ingress and egress interface, denoted by $i$ and $e$.
At time $t$, an ARP request arrives at the ingress interface of port 2.
Then at time $t+1$, the request is at the egress interfaces of ports 3 and 4
as we would expect since port 1 is uplink and the frame is flooded to all ports except its ingress port.
An ARP reply is received at time $t+2$ at port 3 and fowarded
to port 2 at time $t+3$ because its destination address was learned there at time $t$.
Elements of a trace are referenced within a recognizer by free variables
$t$ (current time), $f$ (frame in the trace at time $t$),
$\id{loc}$ (location of $f$) and
$\id{port}$ (the ingress port of $f$ when $f$ is located at an ingress interface).

\subsection{Hub component $H(\id{self})$}

Hub component $H(\id{self})$
is defined in Table~\ref{relay} using a special type of recognizer called a $\lambda$-SFA.
It is a type of deterministic symbolic 
finite automaton (SFA) \cite{vannoord2001, veanes2010} with lambda bindings that
allow it to more succinctly remember 
history.\footnote{
$\lambda$ is an input binding operator as in $\lambda$ calculus, not a name for the null
string as in finite automata.}
\begin{table}[h]
\caption{$H(\id{self})$ relays between non-uplink ports}
\small
\centering
\begin{tabular}{l}
{\bf H1} $\rightarrow$ {\bf H1} \\ $\begin{array}{l}
\id{loc} = \{\id{port}\;\rid{i}\} \Rightarrow (\id{port} = \id{uplink-port} \vee f.\id{da} = \id{haddr}(\id{port}))
\end{array}$  \\

{\bf H1} $\rightarrow$ {\bf H2} \\ 
$\begin{array}{l}
\lambda x.\;\id{loc} =\{\id{port}\;\rid{i}\}\wedge
\id{port}\neq\id{uplink-port}\wedge 
f.\id{da}\neq \id{haddr}(\id{port})
\end{array}$ \\ 

{\bf H2} $\rightarrow$ {\bf H1} \\ 
$\begin{array}{l}
(\id{self}\;\rid{e} \in\id{loc} \wedge ((\id{bcast}(x.f.\id{da})\wedge \neg\id{arp-reqrx}(x.f,x.\id{port}))\; \vee \\
\id{ucast}(x.f.\id{da}))) \Rightarrow 
(f = x.f \wedge \id{self} \neq x.\id{port} \wedge \id{self} \neq \id{uplink-port})
\end{array}$
\end{tabular}
\label{relay}
\end{table}
$H(\id{self})$ has three transitions and two states H1 and H2 where H1 is the start state (the first transition listed
is always from the start state).
The proposition that labels a transition is shown below it.
A transition from H1 to H2 occurs when a frame arrives at an ingress port other than the uplink port and
its destination hardware address $f.\id{da}$ doesn't match the hardware address of the port, 
which indicates link-layer forwarding rather than handling traffic destined for the switch.
Otherwise it stays in H1.
Notice that if $H(\id{self})$ were started in state H1 at time $t+1$, then it stays in H1 because $\id{loc}=\{\rid{3e},\rid{4e}\}$
at that time, and thus $\id{loc}=\{\id{port}\;\rid{i}\}$ is false then.
This is a stutter step that allows the SFA to ignore actions in a trace that are not of interest to it 
in state H1, namely egress activity \cite{TLA1994}.
For the trace in Table~\ref{trace},
the bindings of the free variables of $H(\id{self})$ 
are given in Table~\ref{bindings}.
\begin{table}[h]
\caption{Free variables of $H(\id{self})$ bound by trace in Table~\ref{trace}}
\small
\begin{tabular}{lllll}
{\em time} & $\id{f.da}$ & $\id{x.f.da}$ & {\em port} & $\id{x.port}$ \\
t & ff:ff:ff:ff:ff:ff & $-$ & 2 & $-$  \\
t + 1 & ff:ff:ff:ff:ff:ff & ff:ff:ff:ff:ff:ff & $-$ & 2 \\
t + 2 & 04:0c:ce:d2:08:6c & ff:ff:ff:ff:ff:ff & 3 & $-$ \\
t + 3 & 04:0c:ce:d2:08:6c & 04:0c:ce:d2:08:6c & $-$ & 3 
\end{tabular}
\label{bindings}
\end{table}

With respect to our 4-port switch, $H(\id{self})$ has four recognizer instances $H(1)$$-$$H(4)$, one for each port.
Assuming that the ARP request in the trace is not a request for the hardware address of port 2
($\neg\id{arp-reqrx}(x.f,2)$ is true),
each instance can make a transition on every entry in the trace, albeit for different reasons in some states.
At time $t+3$, for instance, all but $H(2)$ move from H2 to H1 by vacuously satisfying its condition
since only 2e is a member of {\em loc\/}.
But $H(2)$ must satisfy its consequent
($f = x.f \wedge 2 \neq 3 \wedge 2 \neq 1$).
If {\em loc\/} were $\{\rid{2e},\rid{3e}\}$ then while $H(2)$ can transition out of state H2, $H(3)$ cannot.
We say $H(3)$ is ``stuck'' in this case.
If {\em loc\/} were $\{\rid{2e},\rid{4e}\}$ then $H(2)$ and $H(4)$ can both
transition out of H2 as $H(4)$ would also satisfy its consequent 
($f = x.f \wedge 4 \neq 3 \wedge 4 \neq 1$).
The fact that {\em loc\/} doesn't include 4e in the trace suggests the switch learned the port for MAC address 04:0c:ce:d2:08:6c.
That brings us to our second component, namely bridging.

\subsection{Bridging component $B(\id{self})$}

The bridging component is given in Table~\ref{switch}.
It forwards a unicast frame only to the port behind which the unicast destination
address was learned.
Like $H$, it is parameterized on {\em self\/}.
The port behind which a MAC address is learned is stored in MAC learning table {\em mlt\/} and
{\em mto\/} is the MAC learning table timeout governing when table entries expire.
For all $i\in\id{dom}(\id{mlt})$,
$\id{mlt}(i).\id{mac}$ is a MAC address that was last seen as an ingress source address at time $\id{mlt}(i).t$ 
at port $\id{mlt}(i).\id{port}$.
\begin{table}[h]
\caption{$B(\id{self})$ bridges between non-uplink ports}
\small
\centering
\begin{tabular}{l}
{\bf B1} $\rightarrow$ {\bf B1} \\ 
$\begin{array}{l}
\id{loc} = \{\id{port}\;\rid{i}\} \Rightarrow (\id{port} = \id{uplink-port} \vee f.\id{da} = \id{haddr}(\id{port}))
\end{array}$  \\

{\bf B1} $\rightarrow$ {\bf B2} \\ 
$\begin{array}{l}
\id{loc} =\{\id{port}\;\rid{i}\}\wedge
\id{port}\neq\id{uplink-port}\wedge 
f.\id{da}\neq \id{haddr}(\id{port})
\end{array}$ \\ 

{\bf B2} $\rightarrow$ {\bf B1} \\
$\begin{array}{l}
(\id{self}\;\rid{e}\in \id{loc} \wedge \id{ucast}(f.\id{da})) \Rightarrow (\\
\dashuline{\exists i.\,\id{mlt}(i).\id{mac} = f.\id{da}\, \wedge \,
t - \id{mlt}(i).t \leq \id{mto} \wedge \id{mlt}(i).\id{port} = \id{self}} \;\vee \\
\underline{\forall i.\,\id{mlt}(i).\id{mac} \neq f.\id{da} \vee t - \id{mlt}(i).t > \id{mto}}\hspace{0.15em})
\end{array}$ 
\end{tabular}
\label{switch}
\end{table}
From state B2, a unicast frame can exit port {\em self\/} only if
the frame's destination address has an entry in {\em mlt\/}, the entry is unexpired and
the port at which the destination address was learned matches the egress port
(dash underlined condition), or the port for the
destination address is unknown or expired (underlined condition).
The latter condition allows a unicast frame to be flooded.

\subsection{Learning component $M$}

The MAC learning table is managed by the learning component defined in Table~\ref{mlt}.
\begin{table}[h]
\caption{$M$ learns MAC addresses at non uplink ports}
\small
\centering
\begin{tabular}{l} 
{\bf ML} $\rightarrow$ {\bf ML} \\ 
$\begin{array}{l}
\lambda x.\,
\dashuline{[(\id{loc} = \{\id{port}\;\rid{i}\} \wedge \id{port} \neq \id{uplink-port} \wedge \id{ucast}(f.sa) \;\wedge} \\
\hspace{1em}\dashuline{(\exists k.\,x.\id{mlt}(k).\id{mac}=f.\id{sa} \;\vee} \;
\dashuline{\exists k.\,t-x.\id{mlt}(k).t > \id{mto}))\; \Rightarrow} \\
\hspace{2em}\dashuline{\exists k.\,\id{mlt}=x.\id{mlt}(k)\,\{\id{mac}=f.\id{sa},t=t,\id{port}=\id{port}\}]} \;\wedge \\

\hspace{1em}\underline{[(\id{loc} \neq \{\id{port}\;\rid{i}\} \;\vee\;}
\underline{\id{port} = \id{uplink-port} \vee \neg\id{ucast}(f.\id{sa}) \;\vee} \\
\hspace{2em}\underline{(\forall k.\,x.\id{mlt}(k).\id{mac}\neq f.\id{sa} \;\wedge\;
\forall k.\,t-x.\id{mlt}(k).t\leq\id{mto})} \\
\hspace{1.75em}\underline{) \Rightarrow \id{mlt} = x.\id{mlt}]}
\end{array}$ 
\end{tabular}
\label{mlt}
\end{table}
It has only one state and merely constrains the MAC learning table in that
either the table is updated (dash underlined condition) or remains unchanged (underlined condition).
An update occurs if a frame arrives at a non-uplink port with a unicast source MAC address and
either that address is already in the table or it's not but there's room in the table for it
because there's an expired entry.
Otherwise the table remains unchanged ($\id{mlt} = x.\id{mlt}$).
It also remains unchanged on egress activity in a trace ($\id{loc}\neq\{\id{port}\;\rid{i}\}$).

We expect a frame to be output in response to every frame input.
The response can be the input frame, a rewrite of it or some other response frame.
This much will be determined by other components, however, we still need a component to enforce
an egress action after every ingress action.
The interleaving component $I$ accomplishes this.
It has an ingress transition $\bid{I1}\rightarrow\bid{I2}$ labeled with 
$\id{loc} = \{\id{port}\;\rid{i}\}$ and an egress transition $\bid{I2}\rightarrow\bid{I1}$ labeled with
$\id{loc} \subseteq \id{egress}$.

\section{Tensor product}	\label{monitor tensor product}

\begin{table}[t]
\caption{$\lambda$-SFA for $H(\id{self})\times B(\id{self})\times I\times M$}
\small
\centering
\begin{tabular}{l}
{\bf H1B1I1ML} $\rightarrow$ {\bf H1B1I2ML} \\ 
$\begin{array}{l}
\lambda x.\;\id{loc} = \{\id{port}\;\rid{i}\} \wedge
\underline{(\id{port} = \id{uplink-port} \vee f.\id{da} = \id{haddr}(\id{port}))} \;\wedge \\
((\id{port} \neq \id{uplink-port} \wedge \id{ucast}(f.\id{sa}) \;\wedge \\
(\exists k.\, x.\id{mlt}(k).\id{mac} = f.\id{sa} \vee
\exists k.\, t- x.\id{mlt}(k).t > \id{mto})) \Rightarrow \\
\hspace{1em}\exists k.\, \id{mlt} = x.\id{mlt}(k)\{\id{mac} = f.\id{sa}, t = t, \id{port} = \id{port}\}) \hspace{2em}\wedge \\
((\id{port} = \id{uplink-port} \vee \neg\id{ucast}(f.\id{sa}) \vee 
(\forall k.\, x.\id{mlt}(k).\id{mac} \neq f.\id{sa} \\
\hspace{1em}\wedge\; \forall k. \, t-x.\id{mlt}(k).t \leq \id{mto} )) \Rightarrow \id{mlt} = x.\id{mlt} )
\end{array}$  \\[4em]

{\bf H1B1I1ML} $\rightarrow$ {\bf H2B2I2ML} \\ 
$\begin{array}{l}
\lambda x.\;\id{loc} = \{\id{port}\;\rid{i}\} \wedge
\dashuline{\id{port} \neq \id{uplink-port} \wedge f.\id{da} \neq \id{haddr}(\id{port})} \;\wedge \\
((\id{ucast}(f.\id{sa}) \;\wedge \\
(\exists k.\, x.\id{mlt}(k).\id{mac} = f.\id{sa} \vee
\exists k.\, t- x.\id{mlt}(k).t > \id{mto})) \Rightarrow \\
\hspace{1em}\exists k.\, \id{mlt} = x.\id{mlt}(k)\{\id{mac} = f.\id{sa}, t = t, \id{port} = \id{port}\}) \hspace{2em}\wedge \\
((\neg\id{ucast}(f.\id{sa}) \vee 
(\forall k., x.\id{mlt}(k).\id{mac} \neq f.\id{sa} \;\wedge \\
\hspace{1em}\forall k. \, t-x.\id{mlt}(k).t \leq \id{mto} )) \Rightarrow \id{mlt} = x.\id{mlt} )
\end{array}$   \\[4em]

{\bf H1B1I2ML} $\rightarrow$ {\bf H1B1I1ML} \\ 
$\begin{array}{l}
\id{loc}\subseteq\id{egress}\hspace{2em}\rid{No action taken for frames destined for switch.}
\end{array}$   \\[1em]

{\bf H2B2I2ML} $\rightarrow$ {\bf H1B1I1ML} \\
$\begin{array}{l}
\id{self}\;\rid{e} \in \id{loc}\Rightarrow [\hspace{0.5em}{\color{blue}(}\\
\hspace{1em}((\neg\id{bcast}(x.f.\id{da}) \vee \id{arp-reqrx}(x.f, x.\id{port})) \wedge 
\neg\id{ucast}(x.f.\id{da})) \;\vee  \\
\hspace{1em}(f = x.f \wedge \id{self} \neq x.\id{port} \wedge \id{self} \neq \id{uplink-port})\hspace{0.5em}{\color{blue})} \;\wedge  \\
\hspace{2em}{\color{red}(}\neg\id{ucast}(f.\id{da}) \; \vee \\
\hspace{1em}(\exists i.\, \id{mlt}(i).\id{mac} = f.\id{da} \wedge t-\id{mlt}(i).t \leq \id{mto} \wedge \id{mlt}(i).\id{port} = \id{self}\\
\hspace{1em}\vee\; \forall j.\, \id{mlt}(j).\id{mac} \neq f.\id{da} \vee t-\id{mlt}(j).t > \id{mto})
{\color{red})}\hspace{0.5em}] \\
\wedge\; \id{loc}\subseteq\id{egress}
\wedge \id{mlt} = x.\id{mlt}
\end{array}$ 
\end{tabular}
\label{HBIM}
\end{table}
Tensor product $H(\id{self})\times B(\id{self}) \times I \times \id{M}$ gives the
semantics of our 4-port switch function and is shown in Table~\ref{HBIM}.
The product is computed with the help of the Yices SMT solver \cite{yices}, 
which eliminates transitions with unsatisfiable propositions.
Notice how the product automatically creates the desired control logic, splitting frame processing into handling frames destined
for the switch (e.g. management frames or frames to be routed), conveyed by the
underlined condition, and those that are not (switched), conveyed by the dash underlined condition.
This happens because interleaving component $I$ doesn't allow the hub component to spin 
on successive ingress frames arriving at the uplink port and remain in state H1.
In state H1I2, $\id{loc}\subseteq\id{egress}$ is true which makes constraint 
$\id{loc} = \{\id{port}\;\rid{i}\} \Rightarrow (\id{port} = \id{uplink-port} \vee f.\id{da} = \id{haddr}(\id{port}))$
on $\rid{H1}\rightarrow \rid{H1}$ vacuously true.
Handling frames destined for the switch function occurs in state H1B1I2ML, 
which is incomplete with respect to the components presented because none of them is concerned
with handling such frames.
Thus this state merely requires $\id{loc}\subseteq\id{egress}$ to transition out.

The switch function between non-uplink ports, on the other hand, is complete.
On the ingress side (state H1B1I1ML), a frame arriving at a non uplink port that is not destined
for the hardware address of the port causes an update to the MAC learning table if its source hardware
address is unicast.
If the source address is already in the table or there's an expired entry allowing it to be inserted
then the learned port and timestamp fields are reset.
If for some reason the source address is not unicast or it is but it's not already in the table and
no entries in the table are expired then the MAC learning table remains unchanged.
On the egress side (state H2B2I2ML), we have {\em self\/} among the egress ports for the output
frame if the input frame is a broadcast but not an ARP request for the ingress
port's hardware address or it's a unicast.
In this case, the current frame $f$ to be output is constrained to be $x.f$ and
{\em self\/} cannot be the uplink port or the ingress port ($x.\id{port}$).
In addition, if the destination hardware address of $f$, which is the destination 
address of $\id{x}.\id{f}$ since $f=\id{x}.\id{f}$,
is unicast then {\em self\/} is governed by the learning component.

\section{Code generation for DPDK platform}

Every formula governing a transition in a product is converted into a minimum disjunctive normal form (DNF).
Branching logic is then computed for each DNF formula.
Finally, the disjuncts of these formulas are discharged into C code using the DPDK API.

Ideally, both branch size and expected running time should be optimized but this isn't always possible.
Minimizing expected running time requires minimizing expected residuals:

\begin{definition}
Given a predicate $p$ and a set of disjuncts $D$, let $\id{res}(p,D)=\emptyset$
if $p\in D$.
Otherwise, a predicate $q$ is in $\id{res}(p,D)$ if $\neg q\not\in\id{res}(p,D)$, $q\neq p$ and there's
a disjunct $d\in D$ such that $q$ occurs in $d$ and $d\wedge p$ is satisfiable.
If $A$ is a truth assignment for members of $\id{res}(p,D)$
then the expected residual of $p$ relative to $D$ and $A$ is
\[
\rid{Pr}[p\,|\,A] \times
|\id{res}(p,D)| +
(1 - \rid{Pr}[p\,|\,A]) \times
|\id{res}(\neg p,D)|
\]
\end{definition}
For instance, consider DNF formula $(C\wedge B)\vee (F\wedge B)\vee E$, so $D=\{C\wedge B, F\wedge B, E\}$.
Suppose predicate $B$ is more likely to be true than $E$, $C$ and $F$, reflected say
by the distribution
$\rid{Pr}[B]=12/16$,
$\rid{Pr}[C]=2/16$ and
$\rid{Pr}[F]=
\rid{Pr}[E]=1/16$.
If $B$ is true then $C$, $F$ and $E$ remain to be evaluated, thus $|\id{res}(B,D)|=3$.
And if it's false then only $E$ remains, so $|\id{res}(\neg B,D)|=1$.
Residuals can likewise be computed for the other predicates.
The expected residuals of the predicates then with respect to $D$
and $A=\emptyset$ become:
\[
\begin{array}{l}
\rid{Pr}[B] \times 3 + (1 - \rid{Pr}[B]) \times 1 = 36/16 + 4/16 = 40/16 \\
\rid{Pr}[C] \times 3 + (1 - \rid{Pr}[C]) \times 3 = 6/16 + 42/16 = 48/16 \\
\rid{Pr}[F] \times 3 + (1 - \rid{Pr}[F]) \times 3 = 3/16 + 45/16 = 48/16 \\
\rid{Pr}[E] \times 0 + (1 - \rid{Pr}[E]) \times 3 = 0 + 45/16 = 45/16
\end{array}
\]
Since $B$ has the least expected residual,
branching would begin by evaluating $B$ to minimize expected running time.
Note that by starting this way, the final branch size will not be minimal
since the minimum size is achieved by evaluating $E$ first.
So it is not always possible to minimize both size and expected running time.

Residual calculations are then made for $D=\{C, F, E\}$ for the ``then'' branch and for $D=\{E\}$ for
the ``else'' branch, each with respect
to $A=\{B\}$.
If $B$ is a predicate asserting a frame is a broadcast, for instance, and $C$ is a predicate
asserting the frame is an ARP request then $\rid{Pr}[C|A]$ is the probability the frame is
an ARP request given it's a broadcast. 
This can vary depending on the network environment of the UPF.
An advantage of our approach is that branching logic can be regenerated continuously in response
to observed traffic that causes the distribution to change.
So the UPF can adapt in real time and remain optimal for the given environment.

After branching is computed for each DNF formula,
the formula's disjuncts are discharged in the context of declarations provided by 
a service-discipline wrapper.
This requires distinguishing checkable predicates from enforceable ones.
The former translates into guards
and the latter into statements of the generated C code.
An {\em enforceable\/} predicate is one whose truth can always be guaranteed at run time,
otherwise, it is {\em checkable\/}.
For example, the formula on the transition from H2B2I2ML in Table~\ref{HBIM} has disjunct:
\[
\begin{array}{l}
\id{self}\;\rid{e}\in\id{loc} \wedge 
\underline{\id{ucast}(\id{x}.\id{f}.\id{da})} \wedge 
f = \id{x}.\id{f} \;\wedge \\
\underline{\id{self}\neq x.\id{port} \wedge \id{self}\neq\id{uplink-port}} \;\wedge \\
\underline{\exists i.\, \id{mlt}(i).\id{mac} = \id{f}.\id{da} \wedge t-\id{mlt}(i).t\leq\id{mto} \wedge 
\id{mlt}(i).\id{port} = \id{self}} \;\wedge \\
\id{loc}\subseteq\id{egress} \wedge \id{mlt}=x.\id{mlt}
\end{array}
\]
Underlined predicates are checkable and all others are enforceable.
Our wrapper code within which generated code runs always guarantees $\id{loc}\subseteq\id{egress}$,
so this predicate can be eliminated at compile time.
Further, the wrapper code runs on a single Intel core so there's no way for a concurrent thread
to change the MAC learning table before entering state H2B2I2ML.
Thus $\id{mlt}=x.\id{mlt}$ can be eliminated (no locking required at run time).
Both predicates are enforceable.
In contrast, the existential constraint on {\em mlt\/} is checkable.
On the surface, there's nothing to suggest it cannot be enforced by an implementation that sets the fields of {\em mlt\/}
as prescribed.
But this cannot be done as it implies control over network function inputs!
$M$ has a single state invariant $\Phi_{\rid{\scriptsize ML}}$ given in
Table~\ref{state invariants}.
\begin{table*}[t]
\caption{State invariants $\Phi_{\rid{\scriptsize B1}}$ and $\Phi_{\rid{\scriptsize ML}}$.}
\small
\centering
\begin{tabular}{l} 
$\Phi_{\rid{\scriptsize B1}}(w_0\cdots w_n,\tau_0\cdots\tau_n)$ : \\[0.25em]
\hspace{0.5em}$\forall i.\,0\leq i< n.$ \\
\hspace{1em}$(\;
((w_i,\tau_i)\models \id{loc} = \{\id{port}\;\rid{i}\}\wedge
\id{port}\neq\id{uplink-port}\wedge 
\id{f}.\id{da}\neq \id{haddr}(\id{port}))\;\wedge\;
((w_{i+1},\tau_{i+1}) \models \id{self}\;\rid{e}\in \id{loc} \wedge 
\id{ucast}(\id{f}.\id{da}))\hspace{0.25em}) \Rightarrow $ \\
\hspace{1.5em}$(w_{i+1},\tau_{i+1}) \models (
\exists i.\,\id{mlt}(i).\id{mac} = f.\id{da} \wedge
\tau_{i+1} - \id{mlt}(i).\id{t} \leq \id{mto} \wedge \id{mlt}(i).\id{port} = \id{self} \;\vee\;$ \\
\hspace{7.5em}$\forall i.\,\id{mlt}(i).\id{mac} \neq f.\id{da} \vee t - \id{mlt}(i).t > \id{mto}$ \\
\hspace{7.5em}$)$ \\[0.5em]

$\Phi_{\rid{\scriptsize ML}}(w_0\cdots w_n,\tau_0\cdots\tau_n)$ : \\[0.25em]
\hspace{0.5em}$\forall d\in\id{dom}(\id{mlt}).\,\forall m,p.\,\forall k.\,0\leq k \leq n.$ \\
\hspace{1em}$((w_k,\tau_k) \models \id{mlt}(d).\id{mac}=m \wedge 
\tau_k - \id{mlt}(d).t \leq \id{mto} \; \wedge \;
\id{mlt}(d).\id{port} = p\hspace{0.25em}) \Leftrightarrow$ \\

\hspace{1.5em}$(w_0\cdots w_k,\tau_0\cdots\tau_k) \models
\exists j.\, 0 \leq j < k.\,\hspace{0.25em}($ \\ 
\hspace{3em}$(w_j,\tau_j) \models (\id{loc} = \{p\;\rid{i}\} \wedge \id{f}.\id{sa} = m \wedge
\id{ucast}(\id{f}.\id{sa}) \wedge 
\exists i.\,\tau_j - \id{mlt}(i).t > \id{mto} \vee \id{mlt}(i).\id{mac} = f.\id{sa}) \;\wedge$ \\
\hspace{3em}$(w_k,\tau_k) \models \tau_k - \tau_j \leq \id{mto} \wedge \id{mlt}(d).t = \tau_j$ \\
\hspace{3em})
\end{tabular}
\label{state invariants}
\end{table*}
It relates the contents of the MAC learning table to an input sequence,
specifically that $\id{f}.\id{da}$ was learned at port {\em self\/} in the past.
This prevents enforcement of the constraint since no implementation can control what is learned at a port.

A discharge table maps predicates to be discharged into C, leveraging
Intel's Data Plane Development Kit (DPDK) \cite{dpdk}.
The DPDK provides a rich API.
For instance, checkable predicate $\id{ucast}(\id{f.da})$ can be discharged directly into C using the DPDK Ethernet API:
\[
\verb+is_unicast_ether_addr(dst_haddr(bufs[buf])) + 
\]
It will be much easier to prove discharge tables correct than to prove entire C programs correct.
Furthermore, it need only be done once.
Thereafter, proving any property of C code generated for a network function will reduce 
to proving properties of finite-state machines ($\lambda$-SFA),
which are easier to reason about than C code.

Our service-discipline wrapper is a simple round-robin service wrapper written in C (580 lines of code)
using the DPDK API (v17.05) \cite{dpdk}
and running on an 8-core Intel Xeon 2.1Ghz server
with 4 X540-AT2 10Gb Ethernet NICs, one for each port of our switch function.
It repeatedly gets for each port a burst of frames using the DPDK API. 
For each ingress frame, it resets the port mask and current time by reading
the timestamp counter register.
It then executes our generated code, producing an output frame and a port mask
defining the egress ports of the frame.
It is a simple service discipline.
Other disciplines like deficit round robin could be used instead.

\section{Proving component properties}	\label{Invariants}

The correctness of a given component is established relative to a requirement formulated
as a property of a timed state sequence \cite{tptl}.
One formulates invariants for the states of the component and proves them by mutual induction.
As examples, we have formulated invariants for
state $\rid{B1}$ of learned forwarding component $B(\id{self})$ and state $\rid{ML}$ of MAC learning component $M$.
They are shown in Table~\ref{state invariants}.
$\Phi_{\rid{\scriptsize B1}}$ relates the current frame to the MAC learning table, and
$\Phi_{\rid{\scriptsize ML}}$ relates the MAC learning table to timed state sequences.
More precisely, $\Phi_{\rid{\scriptsize B1}}$ says if a unicast frame, arriving at a non-uplink port,
is not destined for the switch and at the next time step $\tau_{i+1}$ it exits at port {\em self\/}
then the MAC learning table at time $\tau_{i+1}$ either has an unexpired entry for it, consisting
of its destination MAC address and the port {\em self\/}, or does not.
$\Phi_{\rid{\scriptsize ML}}$ on the other hand states what is true of all destination MAC address/port pairs $(m,p)$
stored in the MAC learning table relative to timed state sequences.
Specifically, destination address $m$ is the source MAC address of a frame that arrived at port $p$ at some
time $\tau_j$ prior to $\tau_k$ where $\tau_k - \tau_j \leq \id{mto}$.

Putting the two invariants together then gives us that $p$ is the port at which destination address $m$ 
was seen as a source MAC address within the last
{\em mto\/} seconds.
Note the invariants alone are insufficient for relating the current frame to a timed state sequence but together
they accomplish it in the product state $\rid{H1B1I1ML}$, which has partial invariant
$\Phi_{\rid{\scriptsize B1}}\wedge\Phi_{\rid{\scriptsize ML}}$.

The invariant of a product state in general is the conjunction of invariants of its component states.
The proof is a straight-forward extension of the standard correctness proof for product automata \cite{Kozen97}.
This homomorphic property is what allows proofs about properties of individual components to scale up to
proofs about properties of products at no extra cost.
This is key to making verification practical for 5G providers.

State invariants are proven by mutual induction on the length of a timed state sequence.
Suppose $\hat{\delta}$ is the multistep transition function for a transition function $\delta$ \cite{Kozen97},
defined as $\hat{\delta}(q,(w_0,\tau_0),\sigma)=(q,\sigma)$ and for $n > 0$,
$\hat{\delta}(q,(w_0 \cdots w_n,\tau_0\cdots\tau_n),\sigma)=(p,\sigma'')$
if
\[
\hat{\delta}(q,(w_0\cdots w_{n-1},\tau_0\cdots \tau_{n-1}),\sigma)=(q',\sigma')
\]
and $\delta(q',(w_n,\tau_n),\sigma')=(p,\sigma'')$.
Note there is no empty timed state sequence; $(w_0,\tau_0)$ reflects the initial state and $\tau_0$ the time
at which initialization of that state is complete.
It forms the base case for induction over sequences.
Then we can show for all 
MAC addresses $m$ and sequences
$\mu = (w_0\;w_1\;\cdots\;w_n,\tau_0\;\tau_1\;\cdots\;\tau_n)$ satisfying
\[
(w_0,\tau_0) \models 
\forall d.\,\tau_0 - \id{mlt}(d).t > \id{mto}  \wedge
\id{mlt}(d).\id{mac}\neq m
\]
if $\sigma$ and $\sigma_0$ are mappings where $\sigma_0(x)=(w_0,\tau_0)$
and $\hat{\delta}(\rid{ML},\mu,\sigma_0) = (\rid{ML},\sigma)$ then
$\Phi_{\rid{\scriptsize ML}}(\mu)$ holds.
Proof is by induction on $n$.

\section{Related work}	\label{related work}

Much work has been done in the design of high-level network programming languages
to configure multiple packet-forwarding devices into a particular network topology \cite{michel2021}.
Frenetic \cite{foster2011}, NDlog \cite{loozhou2012}, OpenBox \cite{openbox},
Nettle \cite{voellmy2011} and P4 \cite{P42014}.
All lack an explicit treatement of time and the ability to reason about timeouts.
In \cite{dobrescu2014}, the aim is to verify bounded execution and crash freedom
for dataplanes constructed as a packet processing pipeline of Click elements that
do not share mutable state beyond the packet and its metadata.
The efforts of \cite{liu2018,neves2018} involve annotating P4 dataplane code with 
assertions and looking for an initial state that leads to their violation.
None of this work can reason about time, history or mutable state.
OpenBox is unique in that it attempts to define the intersection of packet-processing pipelines
via a merge algorithm on packet processing graphs.
However the algorithm is described informally so its soundness is difficult to assess, especially
with potential packet modification conflicts.

An intermediate network program representation, called a network transaction automaton, is described
in \cite{haoli2020,haoli2020b}.
However the product of such automata is not well defined.
A transition can assign to a variable and the product construction requires taking the union of two assignments.
But what is the union of $x:=0$ and $x:=1$?
NetKat \cite{netkat} allows one to specify forwarding policies via a small set of primitive commands
and combinators.
NetKat expressions can be represented as deterministic finite automata (DFA).
So the intersection of policies is defined by the standard product of DFA,
which is an instance of a tensor product.
Temporal NetKat \cite{temporalnetkat}, NetKat extended with linear temporal operators, also lacks
an explicit treatment of time.
 
Emphasis on reusability can be found in the early work around kernel network stack
development: $x$-kernel \cite{hutchinson1991}, Scout \cite{mosberger1996, peterson1999},
and later in extensible routers \cite{decasper1998, click2000, keller2002} and 
decomposition of security services in SDN networks \cite{fresco2013}.
Click \cite{click2000}, is a Linux-based platform
for building a single network stack from reusable C++ classes or ``elements'' linked together to form a
packet-processing chain.
An element can be an arbitrarily-complex computation though in practice it usually implements some
basic step in a network stack like fetching a route or decrementing a TTL.
The work does not facilitate rigorous construction of network functions from reusable parts.
Although packet-processing functions may be reusable they are not expressed in a way that is well suited 
for combining them algorithmically.
In Click, they are C++ programs.

On the formal verification front,
work has been done verifying controllers of software-defined networks (SDN) and dataplanes.
A compiler and run-time system for NetCore \cite{netcore} is verified with mechanical support in \cite{gurefo2013}.
NICE \cite{nice2012}, FlowLog \cite{flowlog2014}, Kuai \cite{kuai2014} and Kinetic \cite{kinetic2015} use model checking
to verify temporal and nontemporal properties of applications like MAC address learning.
Vericon \cite{vericon} takes a different approach, formulating invariants of networks and
properties of SDN programs in first-order logic and then checking satisfiability using Z3.

In \cite{dobrescu2014}, the aim is to verify bounded execution and crash freedom
for dataplanes constructed as a packet processing pipeline of Click elements that
do not share mutable state beyond the packet and its metadata.
The efforts of \cite{liu2018,neves2018} involve annotating P4 dataplane code with 
assertions and looking for an initial state that leads to their violation.
None of this work can reason about time, history or mutable state.

Zen is a modeling language that allows one to express and analyze a wide variety of network functions
written in C\# \cite{zen2020}.
Composing two Zen models is purely operational in that a function of one model can
call a function of the other.
No attempt is made to define it denotationally, for instance,
in terms of a new a property exhibited by the composition that a programmer can inspect.
A declarative language limited to application-layer gateway processing is given in \cite{balldin20}.
Using Z3 one can verify the correctness of packet filtering and rewrite rules.

\section{Conclusions}	\label{conclusions}

Rather than writing networking software and then reasoning about it, the approach presented here involves
generating code from products of primitive reusable components that capture various 
network behaviors.
An example product was given with four components.
These can be mechanically combined to produce a new functional specification from which code is
ultimately generated.
It is easy to add other components that introduce new features like per-port stateful firewalling,
network address translation and so on.

No ex post facto reasoning about generated code is needed once discharge tables are proved correct.
Generated code is not modified directly since changes
are made at the reusable component level.
Consequently, opportunities for introducing low-level bugs in C are eliminated.
Contrast this with the state of the art where bugs can be introduced and then
code must be analyzed to detect them.
If such analysis requires one to annotate dataplane code with assertions and then check whether
the code is a model of them then why bother write the code at all?
Instead one should focus on the assertion logic and derive code from it, making model checking unnecessary.
Others are reaching the same conclusion for SDN controller software \cite{mcclurg2018}.
The challenge then shifts from verifying code to compiling logical assertions into code that rivals
the performance of handwritten dataplane code, a challenging but more tractable problem.

\bibliographystyle{ACM-Reference-Format}
\bibliography{net-functions}

\end{document}